\documentclass[12pt]{article}
\pdfoutput=1
\usepackage{epsfig,amsfonts,amsthm}
\usepackage{amsmath,amssymb}
\newcommand{\be}{\begin{equation}}
\newcommand{\ee}{\end{equation}}
\newcommand{\bea}{\begin{eqnarray}}
\newcommand{\eea}{\end{eqnarray}}

\renewcommand{\Re}{\mathrm{Re }}
\renewcommand{\Im}{\mathrm{Im }}

\newcommand{\Z}{\mathbb{Z}}

\providecommand{\RR}{{\mathbb{R}}}

\def\lsim{\mathrel{\rlap{\lower4pt\hbox{\hskip1pt$\sim$}}
    \raise1pt\hbox{$<$}}}         
\def\gsim{\mathrel{\rlap{\lower4pt\hbox{\hskip1pt$\sim$}}
    \raise1pt\hbox{$>$}}}         

\title{Geometric minimization of highly symmetric potentials}

\author{A.~Degee$^{1}$, I.~P.~Ivanov$^{1,2}$, V.~Keus$^{1,3}$  
\\
  {\small $^1$ IFPA, Universit\'{e} de Li\`{e}ge, All\'{e}e du 6 Ao\^{u}t 17, b\^{a}timent B5a, 4000 Li\`{e}ge, Belgium}\\
  {\small $^2$ Sobolev Institute of Mathematics, Koptyug avenue 4, 630090, Novosibirsk, Russia}\\
  {\small $^3$ School of Physics and Astronomy, University of Southampton, Southampton SO17 1BJ, UK}
  }

\begin{document}
\maketitle

\begin{abstract}
In non-minimal Higgs mechanisms, one often needs to minimize highly symmetric Higgs potentials.
Here we propose a geometric way of doing it, 
which, surprisingly, is often much more efficient than the usual method.
By construction, it gives the global minimum for any set of free parameters of the potential,
thus offering an intuitive understanding of how they affect the vacuum expectation values.
For illustration, we apply this method to the $S_4$ and $A_4$-symmetric three-Higgs-doublet models.
We find that at least three recent phenomenological analyses of the $A_4$-symmetric model
used a local, not the global minimum.
We discuss coexistence of minima of different types, 
and comment on the mathematical origin of geometrical $CP$-violation
and on a new symmetry linking different minima.
\end{abstract}

\section{Introduction} \label{section-introduction}

Finding out the nature of the electroweak symmetry breaking (EWSB) is one of the hottest topics 
in high-energy physics these days.
The first LHC data on the Higgs-like resonance at 126 GeV
show intriguing deviations from the Standard Model (SM) expectations, \cite{discovery},
and many believe that they hint at a non-minimal Higgs mechanism of EWSB.

In the past decades, many non-minimal Higgs sectors have been considered, \cite{CPNSh}.
Typically, these sectors involve several Higgs fields interacting via the scalar potential, 
which is often invariant under a group of Higgs-family transformations. 
Once the potential is written, one then proceeds by minimizing the potential, 
finding the vacuum expectation value (vev) alignment,
expanding the potential near this point, and calculating phenomenologically relevant quantities.

The standard procedure for minimization of the potential is to parametrize the vev's via (possibly complex) $v_i$,
calculate $V(v_i)$, then set all $\partial V/\partial v_i = 0$, solve these equations for $v_i$, and finally 
check that the hessian at this point is positive definite.
Alternatively, one can start with the desired vacuum configuration and build the potential
with a prescribed symmetry around the vacuum point.

Although this method usually works well, there are several reasons why one might not be completely satisfied with it.
First, sometimes the equations cannot be solved analytically.
Second, if they are solvable, they do not always give a clear intuitive picture of
how the vev alignment depends on the free parameters of the potential.
Third, the potentials can support several local minima, and in order to be sure that 
one works with the global minimum, one must check that all other possible minima
of the given potential lie above the chosen one.
Unfortunately, this check is done very rarely, merely because this is a difficult task on its own.

In this paper we propose a geometric method of minimization of potentials which is free from these drawbacks.
Its advantages are:
\begin{itemize}
\item
by construction, it gives the global minimum of the potential,
\item
one does not need to repeat the calculations for various regions of free parameters;
once the main geometrical object (the orbit space) is constructed, one gets answers
for all allowed free parameters,
\item
even in cases when the analytic minimization is impossible, it can still give useful information
e.g. number of degenerate minima, symmetry breaking properties, etc.
\end{itemize}
The main drawback of this method is that it is handy only for highly symmetric potentials,
so that the number of free parameters of the potential is small.
However this situation takes place in many particular realizations of non-minimal Higgs sectors,
and this method should indeed be useful in practical calculations.

In this paper, we will illustrate this method with the particular versions of the three-Higgs-doublet model (3HDM),
the ones with the $A_4$ and $S_4$ Higgs-family symmetry group.
This models has been actively studied in the past few years
\cite{A43HDM,lavouraA4,MorisiPeinadoA4,MachadoA4,toorop1,toorop2}, with the idea that
it might provide a natural explanation to the patterns observed in the fermion mass matrices.
Minimization of the potential was conducted in these papers in the standard way, and several
vev alignments were used.
However, we will show below that some of them correspond in fact to a local, not a global minimum;
a fact that apparently went unnoticed up to now.

The structure of the paper is as follows. 
In Section \ref{section-method} we describe our main idea, which we formulate for convenience in the context of multi-Higgs-doublet models.
We illustrate the general construction in Section \ref{section-3HDM}, where we discuss in detail the $S_4$ and $A_4$-symmetric
3HDM. In Section \ref{section-conclusions} we comment on previous publications,
discuss various additional aspects of the method, and finally draw our conclusions.
In Appendix, for completeness, we present the Higgs mass spectra for all global minima possible
in $S_4$ and $A_4$-symmetric 3HDM.

\section{Geometric minimization of symmetric potentials}\label{section-method}

Although the method we propose is rather general and can be applied to a broad range
of extended Higgs sectors and perhaps beyond, we prefer to expose it in the context of $N$-Higgs-doublet models (NHDM).
This will allow us to keep the notation simple and, at the same time, get prepared for the particular
applications in 3HDM. 

\subsection{Orbit space in NHDM}

In the $N$-Higgs-doublet model we introduce $N$ Higgs doublets $\phi_a$, $a=1,\dots,N$ 
with identical electroweak quantum numbers.
The general renormalizable Higgs potential of NHDM is constructed from the gauge-invariant
bilinear combinations $(\phi_a^\dagger \phi_b)$ \cite{bilinears-original}, 
which describe the gauge orbits in the Higgs space
\footnote{Strictly speaking, $\phi_a$ are operators acting on the Higgs Fock space; 
however, for the purposes of this paper one can view them as doublets of complex numbers.}. 
The space of gauge orbits (the orbit space) can be represented as a certain algebraic manifold
in the space of these bilinears.
It is convenient to group them in the following $N^2$ real bilinears:
\be
r_0 = \sqrt{{N-1\over 2N}}\sum_a \phi_a^\dagger \phi_a\,,\quad r_i = \sum_{a,b} \phi_a^\dagger \lambda^i_{ab}\phi_b\,,
\quad i = 1, \dots, N^2-1\,.
\label{rmu}
\ee
where $\lambda^i$ are the generators of $SU(N)$.
The orbit space as a manifold in the euclidean space $\mathbb{R}^{N^2}$ of these bilinears
was characterized algebraically and geometrically in \cite{ivanovNHDM}.
It lies between two forward cones defined by 
\be
r_0 \ge 0\,,\quad {N-2 \over 2(N-1)}r_0^2 \le \vec r^2  \le r_0^2\,.
\ee
In addition, neutral vacua always lie on the surface of the outer cone $\vec r^2 = r_0^2$,
while charge-breaking vacua lie strictly inside, $\vec r^2 < r_0^2$. 

In the formalism of bilinears, the Higgs potential takes the form of a general quadratic form of $r_0$ and $r_i$:
\be
\label{potential}
V = - M_0 r_0 - M_i r_i + {1 \over 2}\Lambda_{00} r_0^2 + \Lambda_{0i} r_0 r_i + {1 \over 2}\Lambda_{ij} r_i r_j\,.
\ee
The minimization of the potential can then be cast into a geometric condition of contact of two algebraic surfaces constructed
in $\mathbb{R}^{N^2}$: the orbit space and the equipotential surfaces defined by $V = const$, see details in \cite{ivanovNHDM2}.

In the particular case of 3HDM, the bilinears are
\bea
&& r_0 = {(\phi_1^\dagger\phi_1) + (\phi_2^\dagger\phi_2) + (\phi_3^\dagger\phi_3)\over\sqrt{3}}\,,\ 
r_3 = {(\phi_1^\dagger\phi_1) - (\phi_2^\dagger\phi_2) \over 2}\,,\ 
r_8 = {(\phi_1^\dagger\phi_1) + (\phi_2^\dagger\phi_2) - 2(\phi_3^\dagger\phi_3) \over 2\sqrt{3}} \quad
\nonumber\\
&&r_1 = \Re(\phi_1^\dagger\phi_2)\,,\quad 
r_4 = \Re(\phi_3^\dagger\phi_1)\,,\quad 
r_6 = \Re(\phi_2^\dagger\phi_3)\,,\nonumber\\[2mm] 
&&r_2 = \Im(\phi_1^\dagger\phi_2)\,,\quad
r_5 = \Im(\phi_3^\dagger\phi_1)\,,\quad
r_7 = \Im(\phi_2^\dagger\phi_3)\,. \label{ri3HDM}
\eea
The orbit space in 3HDM is defined by 
\be
r_0 \ge 0\,,\quad \vec r^2 \le r_0^2\,,\quad \sqrt{3}d_{ijk} r_i r_j r_k = {3 \vec r^2 - r_0^2\over 2}r_0\,,
\label{3HDMconditions}
\ee
and the modulus of the vector $\vec r$ is restricted as
\be  
{1 \over 4} r_0^2\le \vec r^2 \le r_0^2\,.\label{3HDM_n}
\ee

\subsection{Geometric minimization: the main idea}\label{section-minimization-generic}

The crucial feature of passing from fields to bilinears is that the Higgs potential is simplified;
it becomes a quadratic form of these new variables.
This transition was used in the two-Higgs-doublet model (2HDM) \cite{bilinears}, 
and it allowed one to observe and exploit interesting geometric features 
of the potential both in 2HDM \cite{ivanov2HDMgeometry} and in multi-Higgs-doublet models \cite{ivanovNHDM2}.
Here we propose to go further in this direction, and introduce new variables
in terms of which the potential becomes a {\em linear} function.

To this end, consider an NHDM potential with a sufficiently high symmetry so that $M_i=0$ in (\ref{potential}).
Absence of this term is a hallmark of so-called ``frustrated symmetries'' in NHDM which were discussed
in \cite{frustrated}. The quartic part of the potential contains $k$ different terms, $k$
usually being rather small for a highly symmetric potential. Let us generically write the potential as
\be
V = - M_0 r_0 + r_0^2\sum_{i=0}^k \Lambda_i x_i\,.\label{potential2}
\ee
Here $x_i$ are the quartic terms divided by $r_0^2$, with $x_0=1$ by convention, 
and $\Lambda_i$ are coefficients in front of them. 

Let us now consider the variables $x_i$, which can always be chosen real.
Calculating them for all possible field configurations (or for all possible values of $r$'s inside the orbit space) 
will fill a certain region in the space $\RR^k$.
This region, which we denote by $\Gamma$, is the orbit space ``squashed'' into the $x_i$ space.
Note that the map from $r$'s to $x_i$ is not, generally speaking, injective
because different $r$'s can correspond to the same point $x_i$.

\begin{figure} [ht]
\centering
\includegraphics[height=4cm]{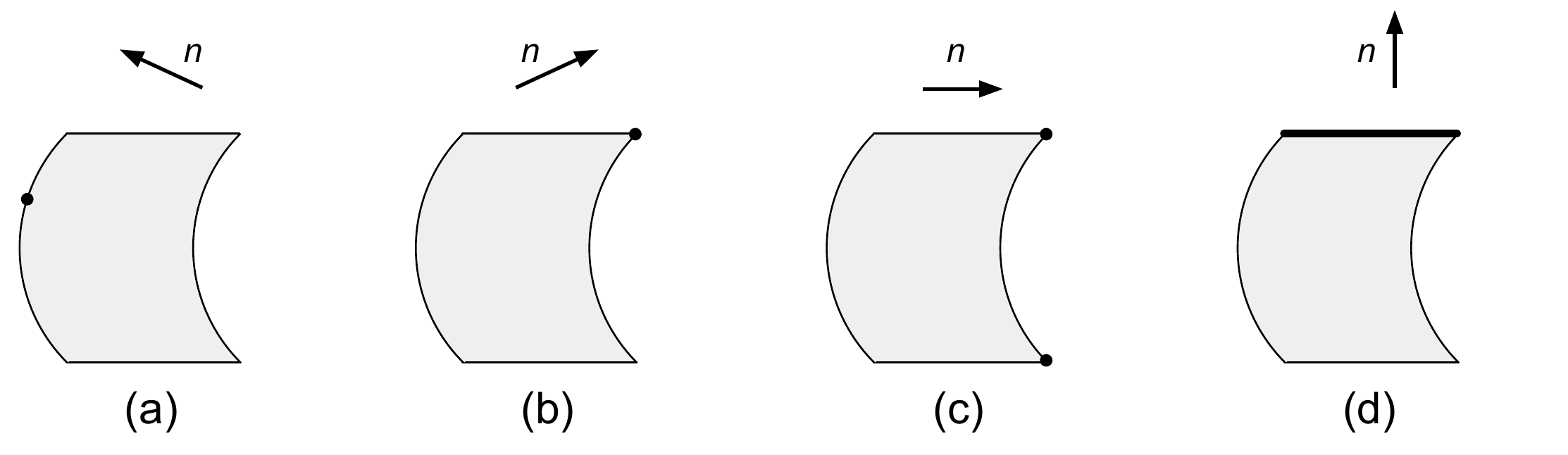}
\caption{Two-dimensional illustration of the geometric minimization method.
Shown are four cases with the same orbit space $\Gamma$ (represented by the shaded region) 
but with different values of the parameters $\Lambda_i$, which define the direction of steepest descent $\vec n$.
The four cases show how the local geometry of $\Gamma$ determines the stablity and degeneracy of the global minimum;
(a): locally convex geometry leads to a single global minimum indicated by the dot, whose position
is sensitive to the exact values of $\Lambda_i$;
(b): a minimum at a vertex is stable against variations of $\Lambda_i$;
(c): cusps separated by a concave region allow two distinct minima to coexist and be degenerate;
(d): straight segments can lead to a continuum of global minima (shown by a thick line) for the special values of the parameters, implying
presence of massless bosons. }
\label{fig-minima}
\end{figure}

Suppose the geometric shape of $\Gamma$ is known.
Then the minimization of the potential proceeds in three simple steps.
First, since the potential (\ref{potential2}) is a linear
function of $x_i$, we can introduce the ``direction of steepest descent'' of the potential,
$\vec n = - (\Lambda_1\,, \dots\,, \Lambda_k)$. 
The potential can then be written as
\be
V = - M_0 r_0 + r_0^2\left(\Lambda_0 - \vec n \vec x\right)\,.\label{potential3}
\ee
Then the minimum of the potential
is achieved at the points of $\Gamma$ which protrude farthest in the direction of $\vec n$.
Once these points $x_i$ are known, we can find their realizations in terms of fields,
and, finally, find the value of $r_0$. Note that the positivity conditions require that $\Lambda_0 - \vec n \vec x > 0$
everywhere in $\Gamma$.

Some phenomenologically relevant properties of the minima follow 
from this geometric picture, which we illustrate in Fig.~\ref{fig-minima} 
with a two-dimensional example. 
If $\Gamma$ has a smooth and strictly convex local shape, Fig.~\ref{fig-minima}a,
then the minimum is unique in the $x_i$ space. It is also seen that the minimum point continuously changes
if parameters $\Lambda_i$ are varied.
If $\Gamma$, instead, has vertices, see Fig.~\ref{fig-minima}b, then the minimum point
becomes stable within certain regions of $\Lambda_i$ variation (or alternatively,
regions of possible directions of $\vec n$). 
Note that such a feature is the origin of geometric $CP$-violation in multi-doublet models, see our discussion
in Sect.~\ref{section-geometric-CP}.
At the borders of these regions,
two concurrent minima become degenerate and coexist, Fig.~\ref{fig-minima}c; crossing this border
causes a first order phase transition between the two vacuum configurations.
Finally, if $\Gamma$ contains straight segments, Fig.~\ref{fig-minima}d, then 
for the borderline parameters $\Lambda_i$ we get a continuum of global minima,
which means that the model contains additional massless scalars.

In this picture, the key object becomes the shape of $\Gamma$ rather than 
the parameters of the potential. Once the symmetry group is fixed
and $\Gamma$ is constructed,
many properties of the potential (points of minimum, their degeneracy and coexistence, 
patterns of symmetry breaking, the phase diagram of the model and phase transitions) 
can be immediately read from its shape.

\section{$A_4$ and $S_4$-symmetric 3HDM}\label{section-3HDM}

\subsection{The potentials}

In this Section we will illustrate how the general method works
with the example of $A_4$ and $S_4$-symmetric 3HDM.

The $A_4$-symmetric 3HDM can be represented by the following potential
\bea
V&=&-\frac{M_0}{\sqrt{3}}\left(\phi_1^{\dagger}\phi_1+\phi_2^{\dagger}\phi_2+\phi_3^{\dagger}\phi_3\right)+\frac{\Lambda_0}{3}\left(\phi_1^{\dagger}\phi_1+\phi_2^{\dagger}\phi_2+\phi_3^{\dagger}\phi_3\right)^2\nonumber\\ 
&&+\frac{\Lambda_3}{3}\left[(\phi_1^{\dagger}\phi_1)^2+(\phi_2^{\dagger}\phi_2)^2+(\phi_3^{\dagger}\phi_3)^2-(\phi_1^{\dagger}\phi_1)(\phi_2^{\dagger}\phi_2)-(\phi_2^{\dagger}\phi_2)(\phi_3^{\dagger}\phi_3)-(\phi_3^{\dagger}\phi_3)(\phi_1^{\dagger}\phi_1)\right]\nonumber\\
&&+\Lambda_1\left[(\Re\phi_1^{\dagger}\phi_2)^2+(\Re\phi_2^{\dagger}\phi_3)^2+(\Re\phi_3^{\dagger}\phi_1)^2\right]\nonumber\\
&&+\Lambda_2\left[(\Im\phi_1^{\dagger}\phi_2)^2+(\Im\phi_2^{\dagger}\phi_3)^2+(\Im\phi_3^{\dagger}\phi_1)^2\right]\nonumber \\
&&+\Lambda_4\left[(\Re\phi_1^{\dagger}\phi_2)(\Im\phi_1^{\dagger}\phi_2)+(\Re\phi_2^{\dagger}\phi_3)(\Im\phi_2^{\dagger}\phi_3)+
(\Re\phi_3^{\dagger}\phi_1)(\Im\phi_3^{\dagger}\phi_1)\right]\,,
\label{Tetrahedralgeneral}
\eea
or, in terms of bilinears,  
\bea
V&=&-M_{0}r_{0}+\Lambda_{0}r_{0}^2+\Lambda_{1}(r_{1}^2+r_{4}^2+r_{6}^2)+\Lambda_{2}(r_{2}^2+r_{5}^2+r_{7}^2)+\Lambda_{3}(r_{3}^2+r_{8}^2)
\nonumber\\
&&+\Lambda_{4}(r_{1}r_{2}+r_{4}r_{5}+r_{6}r_{7})\,.
\label{Tetrahedralgeneral2}
\eea
Here parameters $M_0$ and $\Lambda_i$ are assumed to take generic values.
This potential is symmetric under the full achiral tetrahedral group $T_d$ isomorphic to $A_4 \rtimes \Z_2$ of order 24.
This group is generated by independent sign flips of individual doublets, by cyclic permutations of the three doublets,
as well as by a specific type of generalized-$CP$ transformation
(the $CP$-conjugation combined with exchange of any two doublets). 

An alternative way to parametrize the potential was used in \cite{lavouraA4,MorisiPeinadoA4,toorop1,toorop2}.
Our coefficients are related with the coefficients of the alternative parametrization used in \cite{toorop1} as
\be
-{M_0 \over \sqrt{3}} = \mu^2\,,\quad \Lambda_0 = 3\lambda_1 + \lambda_3 \,,\quad 
\Lambda_3 = - \lambda_3\,,\quad \Lambda_{1,2} = \lambda_4 \pm \lambda_5 \cos\epsilon\,,\quad 
\Lambda_4 = -2\lambda_5 \sin \epsilon\,.
\ee

If $\Lambda_4 = 0$, we get the $S_4$-symmetric 3HDM. In the alternative parametrization, 
this is equivalent to setting $\epsilon = 0$.
The potential becomes now symmetric under the full achiral octahedral group $O_h$ isomorphic to $S_4 \times \Z_2$ of order 48,
which is generated by sign flips of the individual doublets, by their permutations, 
and by the $CP$-conjugation.

Since the classification of the finite realizable symmetry groups of the scalar sector in 3HDM is now known \cite{finite3HDM},
we know that restricting the parameters further will never produce any larger finite symmetry group.
It can only lead to continuous symmetry groups, which are necessarily frustrated and must therefore be spontaneously 
broken and produce massless scalars \cite{frustrated}. We disregard this situation on phenomenological grounds.

\subsection{The orbit space in the $S_4$ case}

Let us start with the more restricted model, the $S_4$-symmetric 3HDM.
Written in terms of bilinears, the potential takes form
\bea
V&=&-M_{0}r_{0}+\Lambda_{0}r_{0}^2+\Lambda_{1}(r_{1}^2+r_{4}^2+r_{6}^2)+\Lambda_{2}(r_{2}^2+r_{5}^2+r_{7}^2)+\Lambda_{3}(r_{3}^2+r_{8}^2)\nonumber\\
&=& -M_{0}r_{0}+r_{0}^2(\Lambda_{0} + \Lambda_{1} x + \Lambda_2 y + \Lambda_3 z)\,,
\label{Octahedralgeneral2}
\eea
with the vector $(x,y,z)$ playing the role of $x_i$. 
The positivity conditions for the potential require that
\be
\Lambda_0 + \Lambda_1 x + \Lambda_2 y + \Lambda_3 z > 0 \label{positivityS4}
\ee
everywhere in the orbit space.
Using the properties of bilinears mentioned in Section~\ref{section-method},
we conclude that the three-dimensional orbit space $\Gamma$ must lie inside the truncated pyramid defined by
\be
x, y, z \geq 0\,, \quad {1 \over 4} \leq x+y+z \leq 1\,. \label{restrictions-xyz}
\ee
In addition, it turns out that $y \le 3/4$. Indeed, $y$ can be rewritten as
\be
y = {3 \over 4}\left[1 - {2 (\kappa_{12} + \kappa_{23} + \kappa_{31}) + Q_{\alpha\beta}Q^*_{\alpha\beta} \over (\rho_1 + \rho_2 + \rho_3)^2} \right]\,,
\label{restriction-y}
\ee
where
\be
\rho_a = (\phi_a^\dagger \phi_a) \ge 0\,,\quad \kappa_{ab} = \rho_a \rho_b - |\phi_a^\dagger \phi_b|^2 \ge 0\,,\quad
Q_{\alpha\beta} = \sum_a \phi_{a}^{\alpha}\phi_{a}^{\beta}\,,
\ee
with $a=1,2,3$ numbering the doublets and $\alpha,\beta = +,0$ denoting the upper and lower components
inside doublets. The largest value of $y$ equal to $3/4$ is attained when, first, all $\kappa_{ab}=0$
which selects the neutral vacuum, and then when the lower components of the doublets sum up as
$\sum (\phi_{a}^{0})^2 = 0$.

\begin{figure} [ht]
\centering
\includegraphics[height=4cm]{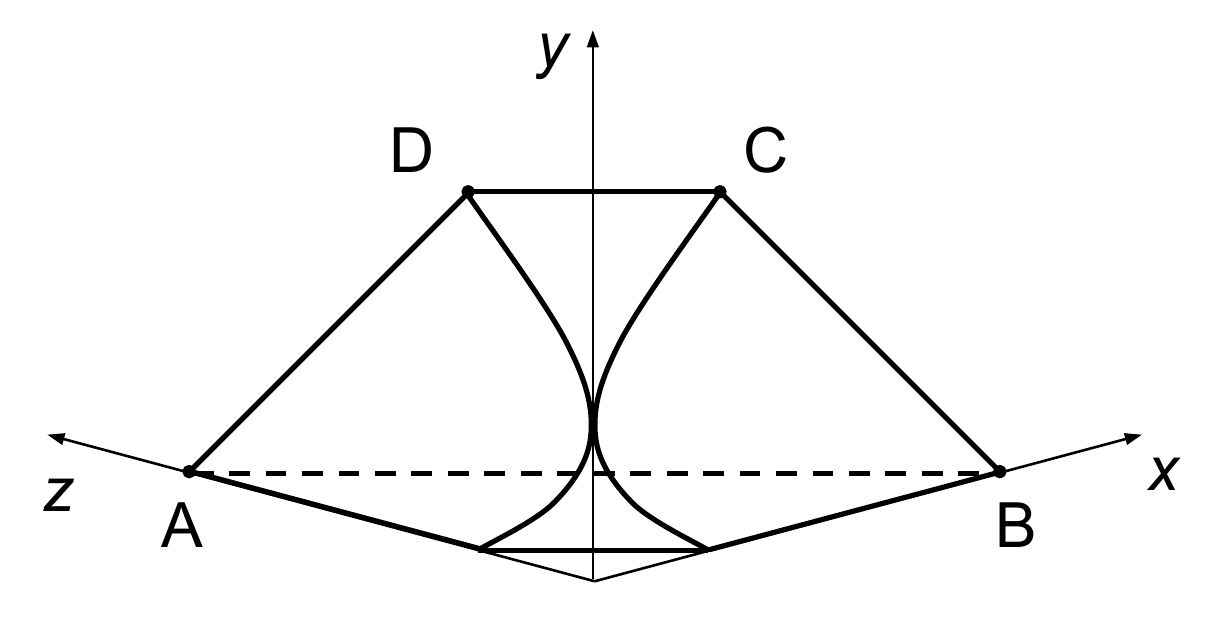}\hspace{5mm}
\includegraphics[height=4cm]{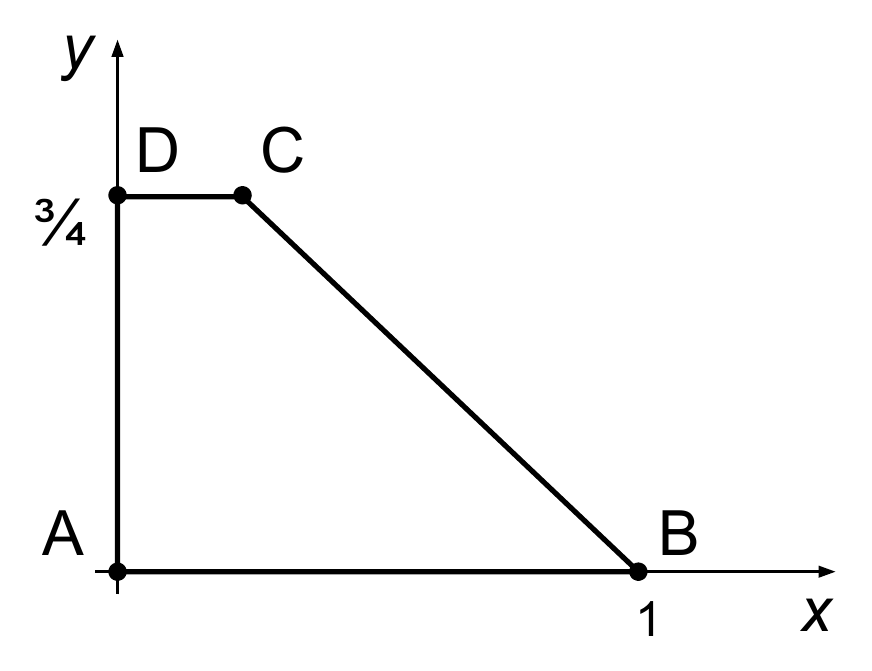}
\caption{Left: sketch of the orbit space $\Gamma$ of the $S_4$-symmetric 3HDM in the $(x,y,z)$-space. Right: the neutral orbit space
in the $(x,y)$-plain. On each plot, the four dots $A$, $B$, $C$, and $D$ mark the positions of the possible neutral global minima.}
\label{fig-xyz3}
\end{figure}

The exact shape of the orbit space which we found by numerical methods%
\footnote{We generated one million points with random up and down components of the three doublets.
For each point we calculated the values of $x$, $y$, and $z$, and then plotted all points. 
By looking at the resulting 3D scatter plot from different angles, we reconstructed the shape and drew the sketch in Fig.~\ref{fig-xyz3}.} is rather complicated, 
see Fig.~\ref{fig-xyz3}, left. It has the form of a wedge with the edge at 
$x+z=1, y=0$ and the convex backside at small values of $x+z$.
However if we focus only on the phenomenologically relevant 
case of neutral vacuum, then we limit ourselves only to one of its faces defined by $x+y+z=1$.
The rest of $\Gamma$ corresponds to charge-breaking vacua and is disregarded.

This neutral part of the orbit space has the shape of a trapezoid, which can be established analytically
using (\ref{restrictions-xyz}) and (\ref{restriction-y}). We show it in Fig.~\ref{fig-xyz3}, right,
on the $(x,y)$-plane (with $z$ defined as $z=1-x-y$).
The vev alignments corresponding to the four vertices of the trapezoid are 
(here we give only the relative magnitude of the vev's)
\be
A:\ (1,0,0)\,,\quad B: \ (1,1,1)\,,\quad C = (e^{{i\pi/3}},e^{-{i\pi/3}},\pm 1)\,,\quad D = (e^{{i\pi/4}},e^{-{i\pi/4}},0)\simeq (1,i,0)\,,
\label{vertices}
\ee
while the straight segments joining them are
\bea
&&AB:\ (v_1,v_2,v_3)\ \mbox{with all $v_i \in \RR$}\,,\quad BC:\ (e^{i\xi_1}, e^{i\xi_2}, e^{i\xi_3})\,,\label{segments}\\
&& AD:\ (v_1, i v_2, 0)\ \mbox{with all $v_i \in \RR$}\,,\quad CD:\ (e^{i\xi}, e^{-i\xi}, r)\ \mbox{with}\ \cos2\xi = -{r^2 \over 2}\,.\nonumber
\eea
Of course, in each case we allow for arbitrary permutations of the doublets.
For example, vertex $D$ corresponds to six degenerate minima $(1,\pm i,0), (1,0,\pm i), (0,1,\pm i)$.

\subsection{Minimization of the $S_4$-symmetric potential}

Applying the methods of Section~\ref{section-minimization-generic}, we immediately conclude that the $S_4$-symmetric 3HDM
can have only four types of neutral minima without producing massless scalars, which correspond to the vertices (\ref{vertices}).
Thus, we located all possible positions of the global minimum without the need to calculate any derivatives.

It is also possible to obtain conditions on $\Lambda_i$ which lead to a minimum at each of these four points
just by looking at the orbit space. 
For example, the vev alignment of the type $(1,0,0)$ becomes the global minimum, when 
$\Lambda_3 < 0$ and $\Lambda_1, \Lambda_2 > \Lambda_3$.
When these conditions are satisfied, the point $A$ indeed lies farthest along the direction $\vec n$.
In addition, the positivity condition (\ref{positivityS4}) in this case implies that $\Lambda_0 + \Lambda_3 > 0$.

\subsection{Unexpected symmetry of the orbit space}\label{section-unexpected-S4}

The $S_4$-symmetric 3HDM possesses a curious feature which could be noticed but would receive no explanation 
with the usual calculations.

First, field content of the scalar sector, after the electroweak symmetry breaking, is the following: apart from the usual three ``would-be'' Goldstone bosons, 
we have two pairs of charge-conjugate Higges $H^\pm_i$, and five neutral scalars.
The oscillation mode in the direction of vev's will be denoted as $h$, while the other neutral Higgses 
are generically labeled as $H_i$. In (generalized) $CP$-conserving cases, these can be additionally classified
as (generalized) $CP$-even and $CP$-odd states.

Let us now calculate the masses of the physical Higgs bosons in the two vev alignments: $(1,1,1)$ and $(1,0,0)$.
In both cases we use $v^2 \equiv v_1^2 + v_2^2 + v_3^2$.
The alignment $(1,1,1)$ becomes the global minimum of the potential if 
\be 
\Lambda_1<0, \quad \Lambda_0 > |\Lambda_1|> -\Lambda_2, -\Lambda_3\,. \label{condition111}
\ee
The minimum point is then parametrized as $(v,v,v)/\sqrt{6}$ with
\bea
&& v^2 ={\sqrt{3}M_0 \over \Lambda_0 - |\Lambda_1|}\,,\label{S4-111}\\[1mm]
m^2_{H^{\pm}_i}&=&  \frac{1}{2}|\Lambda_1|v^2 = \frac{\sqrt{3}M_0}{2}\,{|\Lambda_1| \over \Lambda_0 - |\Lambda_1|} \quad \mbox{(double degenerate)\,,} \nonumber \\[1mm]
m^2_{H_i}&=& \frac{1}{2}(|\Lambda_1|+\Lambda_2)v^2 = \frac{\sqrt{3}M_0}{2}\,{|\Lambda_1| + \Lambda_2 \over \Lambda_0 - |\Lambda_1|} \quad \mbox{(double degenerate)\,,} \nonumber\\[1mm]
&& {1 \over 3}(|\Lambda_1|+\Lambda_3)v^2 = {M_0\over \sqrt{3}}\, {|\Lambda_1| + \Lambda_3 \over \Lambda_0 - |\Lambda_1|}\quad \mbox{(double degenerate)\,,} \nonumber \\[1mm]
m^2_h&=& {2 \over 3}(\Lambda_0-|\Lambda_1|)v^2 = {2 \over\sqrt{3}}M_0\,.
\eea
The alignment $(1,0,0)$ becomes the global minimum if
\be 
\Lambda_3<0, \quad \Lambda_0 > |\Lambda_3|> -\Lambda_2, -\Lambda_1\,. \label{condition100}
\ee
Expanding the potential around the point $(v,0,0)/\sqrt{2}$, we get
\bea
&& v^2= {\sqrt{3}M_0 \over \Lambda_0 - |\Lambda_3| }\,,\label{S4-100}\\[2mm]
m^2_{H^{\pm}_i}&=& \frac{1}{2}|\Lambda_3|v^2 = \frac{\sqrt{3}M_0}{2}\,{|\Lambda_3| \over \Lambda_0 - |\Lambda_3|} \quad \mbox{(double degenerate)\,,} \nonumber \\[2mm]
m^2_{H_i}&=& \frac{1}{2}(|\Lambda_3|+\Lambda_2)v^2 =  \frac{\sqrt{3}M_0}{2}\,{|\Lambda_3| + \Lambda_2 \over \Lambda_0 - |\Lambda_3|}
\quad \mbox{(double degenerate)\,,} \nonumber\\[2mm]
&& \frac{1}{2}(|\Lambda_3|+\Lambda_1)v^2 =  \frac{\sqrt{3}M_0}{2}\,{|\Lambda_3| + \Lambda_1\over \Lambda_0 - |\Lambda_3|}
\quad\mbox{(double degenerate)\,,} \nonumber \\[2mm]
m^2_h&=& {2 \over 3}(\Lambda_0-|\Lambda_3|)v^2 = {2 \over\sqrt{3}}M_0\,.
\eea
It is hard to miss a remarkable symmetry between these mass spectra: upon exchange $\Lambda_1 \leftrightarrow \Lambda_3$
they {\em almost} turn into one another. The only quantity that violates this otherwise perfect symmetry is the mass
of one pair of neutral Higgses. 

The bizarre aspect of this almost perfect symmetry is that it is {\em not a symmetry of the model}.
It would be a symmetry if there existed a transformation of fields that could swap $x$ and $z$
while keeping $y$ unchanged. But such transformation does not exist.
This is also consistent with the fact that $\Lambda_1 \leftrightarrow \Lambda_3$ does not lead to
an exact matching of the two Higgs spectra.

We can trace the origin of this near symmetry from the shape of the orbit space $\Gamma$ in the $(x,y,z)$ space.
Our numerical study offers very strong hints that this shape is indeed $x \leftrightarrow z$ symmetric;
unfortunately, we do not have an analytic proof of this fact.
Provided this is true, it explains why conditions (\ref{condition111}) and (\ref{condition100}) 
and the charged Higgs masses (which are also related with the shape of the orbit space) are exactly symmetric.

It is interesting to notice that, for both vev alignments, the spectrum of the Higgs bosons is {\em 2HDM-like}.
Namely, we have only one value for the charged Higgs masses and three values for neutral Higgs masses,
just as expected for the generic 2HDM. 
How this situations can be distinguished from the true 2HDM experimentally, and which observable quantities 
one should look at, is a separate issue worth investigating further.
However the origin of this 2HDM-like spectra is different in these two cases.
In the vev alignment $(1,1,1)$ it comes form the unbroken $S_3$-symmetry of the model, \cite{A43HDM},
while for the $(1,0,0)$ alignment, the origin is the $O(2)$-symmetry mixing
the second and third doublets, which is manifest in the vev's and the mass terms.
These two symmetry arguments are non-equivalent and cannot be related to each other.
After all, the two vev alignments also differ in the number of degenerate vacuum points: four for the $(1,1,1)$
and three for $(1,0,0)$.
 
The same relation holds between the other two possible vev alignments, see Appendix.
It would be interesting to see if this approximate symmetry 
leads to other phenomenological similarities between these pairs of minima.

\subsection{The orbit space in the $A_4$ case}\label{subsection-tetrahedral}

We write the generic potential of the $A_4$-symmetric 3HDM (\ref{Tetrahedralgeneral2})
in the way suggested in Sec.~\ref{section-minimization-generic}:
\be
V= -M_{0}r_{0}+r_{0}^2(\Lambda_{0} + \Lambda_{1} x + \Lambda_2 y + \Lambda_3 z + \Lambda_4 t)\,,
\label{Tetrahedralgeneral3}
\ee
with the same $x, y, z$ as in Eq.~(\ref{Octahedralgeneral2}) and with $t = (r_1r_2 + r_4r_5 + r_6r_7)/r_0^2$.

\begin{figure} [ht]
\centering
\includegraphics[height=6cm]{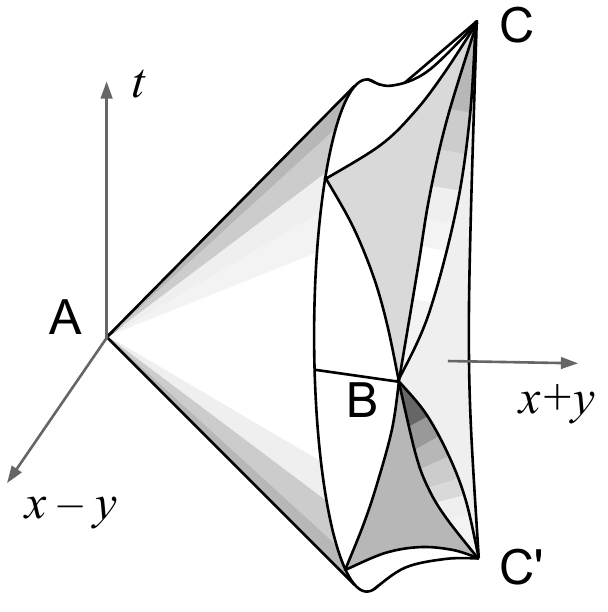}
\hspace{1cm}
\includegraphics[height=6cm]{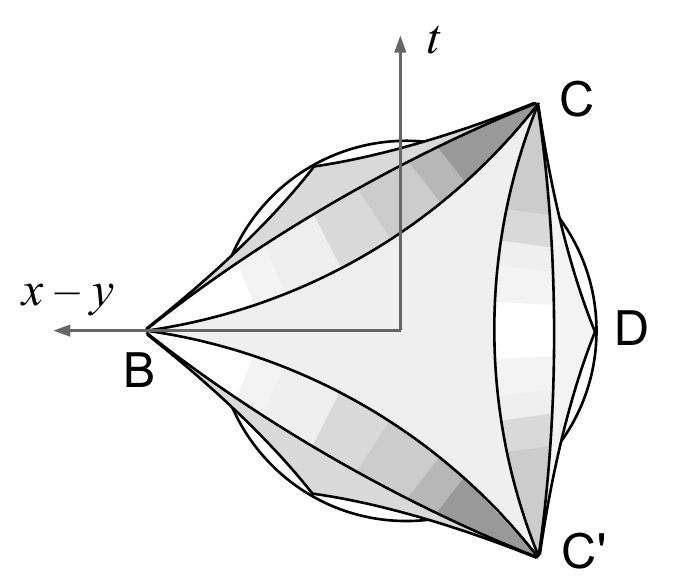}
\caption{Sketch of the neutral orbit space in the tetrahedral 3HDM viewed from two angles.
Uniformly shaded regions correspond to flat faces, graded shading indicates curved faces.}
\label{fig-xyt}
\end{figure}

Again, we focus on the neutral orbit space $\Gamma$, for which we choose $x$, $y$, and $t$
as independent variables, and then $z = 1-x-y$.
The shape of $\Gamma$ which arises from our numerical study is shown in Fig.~\ref{fig-xyt} in the $(x,y,t)$ space.
Despite being rather complicated, it displays a remarkable triangle symmetry.
Basically, it is a right circular cone oriented along the direction $x-y=t=0$ 
with the apex at the origin and with opening angle $\pi/2$
and with directrix lines of length 1. Parts of this cone starting from distance $3/4$ from the apex
are carved out. It has four flat faces which have the shape of deltoid (3-vertex cusped closed curve).
These are located at the bottom face (defined by $z=0$) and at three side faces (one is $y = 3/4$
and the remaining two are obtained by the $2\pi/3$ rotation of the cone).
Directrix lines opposite to these three side faces have length 1 and extend up to the points
$B$, $C$ and $C'$ where three flat deltoid regions meet.
The remaining portions of $\Gamma$ are concave regions.
The two-dimensional neutral orbit space for the $S_4$-symmetric model, Fig.~\ref{fig-xyz3}, right, is simply the 
$(x,y)$-projection of this $\Gamma$;
correspondence between the vertex points in the two shapes should be clear.
We only mention that point $D$, which was a vertex in the $S_4$ case, 
becomes a regular point on the rim of the cone in the $A_4$ case.

The conical shape of the orbit space can be understood in the following way.
Let us introduce two real vectors
\be
\vec a = {1 \over r_0}(r_1,r_4,r_6)\,,\quad 
\vec b = {1 \over r_0}(r_2,r_5,r_7)\,,\quad 
\ee
and denote the angle between them by $\xi$. 
Then, $x = \vec a^2$, $y = \vec b^2$, $t = (\vec a\vec b) = \sqrt{xy}\cos\xi$.
If it is possible for a given point $(x,y)$ to find two parallel vectors $\vec a$ and $\vec b$,
then the orbit space extends in the $t$-direction up to $t = \sqrt{xy}$,
which precisely defines the cone.
It turns out that parallel vectors $\vec a$ and $\vec b$ exist
within the triangle $0 \leq x+y \leq 3/4$ (vev alignment is $(v_1,v_2 e^{i\alpha},0)$), 
and along the straight segments defined by $y=0$ (vev alignment $(v_1,v_2, v_3)$) 
and defined by $y/x = 3$ (alignment $(v_1 e^{i\pi/3}, v_2 e^{-i\pi/3}, v_3)$).

Let us also stress that the emergent triangle symmetry of the orbit space is {\em not}
related to the symmetry of the potential, but is a feature of the orbit space itself.
There is simply no field transformation that realizes rotations of the cone.
In this aspect, this emergent symmetry is similar to the $(x,z)$ reflection symmetry
of the orbit space in the $S_4$-symmetric model.

\subsection{Minimization of the $A_4$-symmetric potential}

The shape of the orbit space immediately leads to the list of possible phenomenologically acceptable global minima 
(i.e. minima not leading to additional massless scalars). These are: the apex of the cone (point $A$),
the three vertices (points $B$, $C$ and $C'$), and the rim of the cone
($x+y=3/4$, $t^2 = xy$). 
Any other point either leads to additional Goldstone bosons or is never a global minimum.

In Appendix, we analyze all these points in some detail and give the Higgs mass spectra.
Comparing these spectra for different minimum points also shows
intriguing relation with the triangle symmetry of the orbit space.

To our knowledge, this is the first complete solution of the minimization problem
in the $A_4$-symmetric 3HDM.

\section{Discussion and conclusions}\label{section-conclusions}

\subsection{Comments on previous publications}

The $A_4$-symmetric 3HDM has received much attention, \cite{A43HDM,lavouraA4,MorisiPeinadoA4,MachadoA4,toorop1,toorop2},
because spontaneous breaking of $A_4$ can generate interesting patterns in fermion mass matrices.
In most papers, the authors just pick up a minimum with specific vev alignment, build a potential
which indeed has a minimum at that point, and then proceed with analysis in the fermion sector. 
In particular, in \cite{lavouraA4,MorisiPeinadoA4} the following vev alignment was used: 
\be
(e^{i\alpha}, e^{-i\alpha}, r)\,, \label{ralpha}
\ee
with $r$ and $\alpha$ being independent real parameters.
Fitting the model to the fermion observables gave $r\sim 40$ in \cite{lavouraA4} and 
$r \sim 240$ in \cite{MorisiPeinadoA4}. 

The authors of \cite{toorop1} aimed at a complete description of possible vev alignments in the minima of the $A_4$-symmetric 3HDM.
They critically reanalyzed 
the phenomenological situation arising at the minimum (\ref{ralpha}) and argued that it is strongly disfavored by
the flavor physics constraints. However, they also considered this vev alignment
as a viable solution of the minimization problem\footnote{We are thankful to Luca Merlo who clarified
to us the motivation behind the works \cite{toorop1,toorop2}.}.

This vev alignment is {\em absent} in our classification because it cannot be the global minimum. 
One can see it most easily precisely in the case of large $r$. Indeed, in our notation,
this point corresponds to $z$ being close to 1 and, therefore, it lies strictly {\em inside the cone} and close to its apex.
This point would emerge at the surface of the cone only for angle $\alpha$ multiple of $\pi/3$,
but in this case one would get massless scalars.
This means that even if this point happens to be a minimum, it is a local, not the global one.	

Thus, we prove that the phenomenological analyses of \cite{lavouraA4,MorisiPeinadoA4} and, partly, of \cite{toorop1}
correspond to a metastable electroweak vacuum. Apparently, this fact went unnoticed because
the authors of these publications did not check whether the potential they got possesses a deeper minimum.

It is generally believed that the Universe sits in the global minimum of the Higgs potential, because
the early hot Universe while cooling down through electroweak temperatures either selected the global minimum
or, even if it were stuck in a metastable vacuum, had enough time to tunnel to the absolute minimum.
Taken seriously, this argument implies that much of the analyses of \cite{lavouraA4,MorisiPeinadoA4}
is not phenomenologically relevant on these grounds, even without appealing to the flavor physics constraints.

\subsection{Relation between the symmetry group and the vacuum structure}

The two examples which we considered in full detail hint at a general relation between the symmetry group
and the vacuum structure: the higher the symmetry group is, the more symmetric is the vacuum alignment.
One facet of this relation is that vev alignments, which cannot be global minima for a very symmetric 
model, might become viable global minima if the symmetry of the potential is lowered.

Indeed, in the $S_4$-symmetric 3HDM, the vev alignment $(1,e^{i\alpha},0)$ with generic $\alpha$ 
is buried deep in the orbit space, and it definitely cannot be the global minimum.
However, in the $A_4$-symmetric 3HDM, this alignment corresponds to the rim of the cone,
and it becomes a viable minimum for certain non-zero $\Lambda_4$.
Geometrically, the extra term in the potential gives an additional dimension to the orbit space,
and in this way it can bring the interior points of the lower-dimensional orbit space 
to the surface of the higher-dimensional orbit space.

Turning again to the vev alignment (\ref{ralpha}) considered in the papers \cite{lavouraA4,MorisiPeinadoA4,toorop1}, 
we can now speculate that even if it is buried deep in the orbit space of the $A_4$-symmetric model,
it might become a viable global minimum in a model with explicitly broken $A_4$. 
In order for this to happen, however, the coefficients in front of the symmetry breaking terms must be sufficiently large. 
Such softly broken $A_4$ models were considered in \cite{toorop2},
a follow-up of \cite{toorop1}. 
Unfortunately, the authors did not check the relative depths of different minima, so that it is not yet know
when (\ref{ralpha}) becomes the global minimum.

\subsection{Origin of geometric $CP$-violation}\label{section-geometric-CP}

The possibility for spontaneous $CP$-violation is one of the motivations behind studying multi-Higgs-doublet models. 
In this context it is often proposed not only that a Higgs-family symmetry should allow
for spontaneous $CP$-violation but also that it should stabilize the vev phases in the global minimum
against variation of the free parameters. 
This situation is known as geometric $CP$ violation, \cite{geometricT,geometricCP,Holthausen:2012dk}
and was originally found in the $\Delta(27)$-symmetric 3HDM 
(though we note that the true Higgs-family symmetry group of that model is $\Delta(54)/\Z_3$, see discussion
in \cite{finite3HDM}).
The relative vev phases arising in geometric $CP$-violation are called calculable 
because their values follow from group theoretic arguments 
and do not depend on the exact values of the parameters of the potential\footnote{However, we 
find the following statement from \cite{geometricCP} inaccurate: ``{\em ...the calculable phase arising from geometrical $CP$ violation 
is uniquely determined independently of the arbitrary parameters of the scalar potential.}''
Indeed, parameters in any case must be such that the point realizing the geometric $CP$ violation 
is the global minimum of the potential. This takes place only in a certain region but not everywhere 
in the space of the free parameters.}.

Using the method described in the present paper, we can pinpoint the mathematical origin
of calculable phases in such models.
They arise due to the presence of vertices in the orbit space $\Gamma$, see Fig.~\ref{fig-minima}b,c,d,
or to be more specific, vertices at points corresponding to non-zero relative phases.
Absence of geometric $CP$ phases would imply convexity of the orbit space, Fig.~\ref{fig-minima}a.  
So, it is not the symmetry of the model {\em per se} that allows for calculable phases
but the choice of coordinates $x_i$ selected by the symmetry, in which the orbit space has vertices.

Our experience shows that the higher the finite symmetry group, the simpler is the geometric shape
of the orbit space $\Gamma$, and the more vertices linked by straight segments it possesses. 
This explains why it is natural that geometric $CP$ violation starts to appear only for sufficiently 
large finite symmetry groups.

\subsection{Coexistence of different minima}

{\em A priori}, it might happen that, for some values of the parameters of the potential, 
two (or more) different types of the global minimum coexist and are degenerate. 
Fig.~\ref{fig-minima}c illustrates this situation.
Upon small variation of the parameters around this special point, one minimum
point becomes the global minimum while the other turns into a local one,
and it is clearly possible to make either of them the global minimum.
This feature leads to a possibility of a first order phase transition upon smooth variation
of the parameters, leading to important phenomenological consequences.
It is therefore desirable to know, which models allow for such a possibility.

It can be inferred from Fig.~\ref{fig-minima}c that this can happen if the
orbit space vertices separated by a concave region, that is, if it has cusps.
If instead the orbit space is a convex body, this possibility is excluded.
Possibility of a first order phase transition is, therefore, linked to the non-convexity 
of the orbit space.

Our analysis shows that the orbit space of the $S_4$-symmetric 3HDM is convex.
Therefore, phenomenologically relevant global minima of different type
cannot coexist in this case.

In the $A_4$-symmetric 3HDM, this possibility arises.
Namely, generic points on the rim of the cone and one of the three vertices $B$, $C$, or $C'$
in Fig.~\ref{fig-xyt} can be degenerate.
It is also possible to make two among these three points degenerate, but not all three.
Examples of such potentials can be readily costructed from geometric analysis of Fig.~\ref{fig-xyt}.

\subsection{How general is the proposed method?}

In which cases does the geometric minimization method proposed in this paper become useful?
Strictly speaking, it has no intrinsic limitation. For example, in the context of the multi-Higgs-doublet models
one can start with an absolutely general Higgs potential, perform a $GL(N,\mathbb{C})$ transformation in the space
of doublets that brings the quadratic term to the form $M_0 r_0$ and then proceeds as discussed in
Section~\ref{section-minimization-generic}. Of course, the potential will contain very many different terms,
so that the orbit space becomes a highly non-trivial multi-dimensional shape.
However comprehending it is only a human limitation, and a hypothetical computer algorithm
could be able to analyze this shape looking for edges, cusps and vertices.

This method becomes much more useful when the number of distinct terms becomes small.
In particular, when the dimension of the neutral orbit space is three or less,
the shape can be relatively easily visualized, and one can develop a much more intuitive picture
of the model than from the usual algebra. For example, in the 3HDM, this situation takes place
for the following finite Higgs-family symmetry groups (based on the classification of \cite{finite3HDM}): 
$A_4$, $S_4$, $\Delta(54)/\Z_3$, and $\Sigma(36)$ (the last two cases not discussed in this paper).
It would be interesting to see if other useful examples appear in other models.

\subsection{Conclusions}

In summary, we have presented a simple yet powerful and very intuitive geometric approach 
to minimization of highly symmetric potentials in non-minimal Higgs mechanisms.
This method is capable of giving the positions of the global minima with very little calculations;
in particular, it avoids the need to differentiate the potential, solve for stationary points, and check the positivity of the hessian.
In a single picture, it shows all points which can be the global minimum for any values of the parameters of the potential.

For illustration purposes, we have applied this method to $A_4$- and $S_4$-symmetric three-Higgs-doublet models
and found all points of global minimum. By doing this, we have also proved that the vacuum point used in at least
three recent phenomenological analyses of the $A_4$ case was not the global, but only the local minimum.
We have also observed an unexpected approximate symmetry linking Higgs spectra at different minima and discussed its origin.
We believe that this method can become a useful tool in all situations 
where minimization of highly symmetric functions is required.

\section*{Acknowledgements} 
It is a pleasure to acknowledge useful discussions with Pedro Ferreira, Lu\`is Lavoura, Rui Santos 
and especially with Jo\~{a}o Silva, who read the draft of this paper and made numerous suggestions. 
I.P.I. also thanks them for hospitality at the University of Lisbon, where this paper was completed.
Communications with L.~Merlo, S.~Morisi and E~Peinado are also acknowledged. 
This work was supported by the Belgian Fund F.R.S.-FNRS,
and in part by grants RFBR 11-02-00242-a, RF President grant for
scientific schools NSc-3802.2012.2, and the
Program of Department of Physics SC RAS and SB RAS "Studies of Higgs boson and exotic particles at LHC."

\appendix

\section{Higgs spectra of the $S_4$-symmetric potential}

In the case of $S_4$-symmetric 3HDM we analyzed the two simple vev alignments in the main text.
They were shown to be approximately related to each other by the unexpected symmetry of orbit space.
The remaining two points also follow this pattern.
The alignment $(1,i,0)$ becomes the global minimum if
\be
\Lambda_2 < 0\,,\quad |\Lambda_2| > |\Lambda_3|\,,\quad \Lambda_1 > \Lambda_3\,,\quad 4\Lambda_0 + \Lambda_3 > 3|\Lambda_2|\,,
\ee
and at this point we have
\be
v^2 ={4\sqrt{3}M_0 \over 4\Lambda_0 + \Lambda_3 - 3|\Lambda_2|}\,,\quad
m^2_{H^{\pm}_i}= {1\over 2}|\Lambda_2|v^2\,, \quad  {1\over 4}(|\Lambda_2| - \Lambda_3) v^2\,,
\ee
and the neutral Higgs masses are
\bea
m^2_{H_i}&=& \frac{1}{4}(\Lambda_1-\Lambda_3)v^2 \quad \mbox{(double degenerate)} \nonumber\\[2mm]
&& \frac{1}{2}(|\Lambda_2|+\Lambda_3)v^2\,,\quad \frac{1}{2}(|\Lambda_2|+\Lambda_1)v^2 \nonumber \\[2mm]
m^2_h&=& \frac{1}{6}(4\Lambda_0 + \Lambda_3 - 3|\Lambda_2| ) v^2 = {2 \over\sqrt{3}}M_0\,.
\eea
The alignment $(\pm 1,e^{i\pi/3},e^{-i\pi/3})$ is the global minimum if 
\be
\Lambda_2 < 0\,,\quad |\Lambda_2| > |\Lambda_1|\,,\quad \Lambda_3 > \Lambda_1\,,\quad 4\Lambda_0 + \Lambda_1 > 3|\Lambda_2|\,,
\ee
and at this point we have 
\be
v^2 = {4\sqrt{3}M_0 \over 4\Lambda_0 + \Lambda_1 - 3|\Lambda_2| }\,,\quad
m^2_{H^{\pm}_i}= \frac{1}{2}|\Lambda_2|v^2,\quad \frac{1}{4}(|\Lambda_2|-\Lambda_1)v^2\,,
\ee
with the neutral Higgs masses
\bea
m^2_{H_i}&=& (a + b \pm \sqrt{a^2 + b^2})\,v^2\quad\mbox{(double degenerate),}
\quad a= \frac{|\Lambda_2| + \Lambda_1}{4}\,,
\quad b = \frac{\Lambda_3 - \Lambda_1}{6}\,,\nonumber\\[2mm]
m^2_h&=&  \frac{4\Lambda_0 +\Lambda_1-3|\Lambda_2| }{6} v^2 = {2 \over\sqrt{3}}M_0\,.
\eea
Again, we observe the perfect $\Lambda_1 \leftrightarrow  \Lambda_3$ symmetry in $v^2$, in the minimum
conditions and in the charged Higgs masses.

\section{Higgs spectra of the $A_4$-symmetric potential}

Let us write for completeness the Higgs mass spectrum at all four possible points
of global minimum found in the main text.
\begin{itemize}
\item
The vev alignment $(1,1,1)$ remains stable in the presence of non-zero $\Lambda_4$ if it is not too large:
\be
\Lambda_4^2 < 12 \Lambda_1 ^2\,,\quad \Lambda_4^2 < 2(\Lambda_3+|\Lambda_1|)(\Lambda_2+|\Lambda_1|)\,.
\ee
The value of $v^2$ is the same as in (\ref{S4-111}), while the masses become
\bea
&&m^2_{H_i^{\pm}}= \left(\frac{1}{2}|\Lambda_1| \pm \frac{1}{4\sqrt{3}}\Lambda_4\right)v^2\,,
\quad m^2_h= {2 \over 3}(\Lambda_0-|\Lambda_1|)v^2\,,\nonumber \\[1mm]
&&m^2_{H_i}= {v^2 \over 12}\left[5|\Lambda_1|+3\Lambda_2+2\Lambda_3 \pm
\sqrt{(|\Lambda_1|+3\Lambda_2-2\Lambda_3)^2 +12\Lambda_4^2}\right]\quad\mbox{(double degenerate)}\nonumber\,.
\eea
Note that the presence of $\Lambda_4$ splits the charged Higgs masses while it preserves
the degeneracy of the neutral Higgses.
\item
The vev alignment $(1,0,0)$ is also stable if $\Lambda_4$ satisfies
\be
\Lambda_4^2 < 4(\Lambda_1+|\Lambda_3|)(\Lambda_2+|\Lambda_3|)\,.
\ee
The value of $v^2$ and the masses of the charged Higgs and the non-degenerate neutral bosons are the same as 
in (\ref{S4-100}),
while the neutral Higgses from the second and third doublets get masses
\be
m^2_{H_i}= {v^2 \over 4}\left[\Lambda_1+\Lambda_2+2|\Lambda_3| \pm
\sqrt{(\Lambda_1-\Lambda_2)^2 +\Lambda_4^2}\right]\quad\mbox{(double degenerate)}\nonumber\,.
\ee
Note that this Higgs spectrum remains 2HDM-like as it was in the $S_4$ case.
\item
The alignment $(1, e^{i\alpha},0)$ for a generic $\alpha$ parametrizes the points around the rim of the cone.
For a given value of $\Lambda_4$, the value of $\alpha$ corresponding to the global minimum is fixed by the relation
\be
\sin 2\alpha = -{\Lambda_4 \over \sqrt{(\Lambda_1 - \Lambda_2)^2 + \Lambda_4^2}}\,,\quad 
\cos 2\alpha = -{\Lambda_1 - \Lambda_2 \over \sqrt{(\Lambda_1 - \Lambda_2)^2 + \Lambda_4^2}}\,.
\ee
Geometrically, this result means that the global minimum lies on the rim in the same direction as the vector $\vec n$
projected on the plane of the rim.
In this case we obtain 
\be
v^2 = {4\sqrt{3}M_0 \over 4\Lambda_0 + \Lambda_3 - 3\tilde \Lambda}\,, 
\ee
where
\be
\tilde\Lambda \equiv - (\Lambda_1 c^2_\alpha + \Lambda_2 s^2_\alpha + \Lambda_4 c_\alpha s_\alpha) 
= {1 \over 2}\left[\sqrt{(\Lambda_1 - \Lambda_2)^2 + \Lambda_4^2} - (\Lambda_1 + \Lambda_2)\right] > 0\,,
\ee
and the mass spectrum is
\bea
m_{H^\pm_i}^2& = & {1 \over 2}v^2 \tilde\Lambda\,,\quad {1 \over 4}v^2 (\tilde \Lambda - \Lambda_3)\,\nonumber\\
m_{H_i}^2 &=&  {1 \over 4}v^2\left[-(\Lambda_3 + \tilde \Lambda) +
 \left(1 \pm \cos3\alpha\right)\sqrt{(\Lambda_1 - \Lambda_2)^2 + \Lambda_4^2}\right]\,,\nonumber\\
&& {1 \over 2} v^2 (\Lambda_1 + \Lambda_2 + 2 \tilde \Lambda)\,, 
\qquad {1 \over 2} v^2(\Lambda_3 + \tilde \Lambda)\,,\nonumber\\
m_h^2 &=& {v^2 \over 6}  (4\Lambda_0 + \Lambda_3 - 3 \tilde \Lambda) = {2 M_0 \over \sqrt{3}}\,.
\eea
Note that presence of $\cos 3\alpha$ in one pair of masses is natural and it reflects the triangle symmetry of the $A_4$ orbit space shown in
Fig.~\ref{fig-xyt}. All other quantities are rotationally invariant, corresponding to the rotational symmetry of the cone.

In the limit $\Lambda_4 \to 0$, we get minimum at $\alpha \to \pi/2$, $\tilde \Lambda \to -\Lambda_2$, 
and these spectra turn into the $S_4$-spectra found in Section~\ref{section-unexpected-S4}.
Note also that the three points on the rim with $\cos 3\alpha = \pm 1$ can never be ``good minima'' because
we get mass terms with coefficients $\pm (\Lambda_3 + \tilde \Lambda)$, which cannot be made positive simultaneously.
These three points lie, in fact, on the three long directrices of the cone.
\item
Finally, in the case of the alignment $(\pm 1, e^{i\alpha}, e^{-i\alpha})$, with $\alpha=\pi/3$, 
we introduce convenient notation
\be
\sin \gamma = {\Lambda_4 \over \sqrt{(\Lambda_1 - \Lambda_2)^2+\Lambda_4^2}}\,, \quad 
\cos \gamma = {\Lambda_1-\Lambda_2 \over \sqrt{(\Lambda_1 - \Lambda_2)^2+\Lambda_4^2}}\,, 
\ee
and
\be
\tilde \Lambda = \Lambda_1 c_\alpha^2 + \Lambda_2 s_\alpha^2 + \Lambda_4 c_\alpha s_\alpha 
= {1\over 2}\left[\Lambda_1 + \Lambda_2 + \cos(2\alpha - \gamma)\sqrt{(\Lambda_1 - \Lambda_2)^2+\Lambda_4^2}\right]\,. 
\ee
Then, the value of $v^2$ is
\be
v^2 = {\sqrt{3} M_0 \over \Lambda_0 + \tilde \Lambda}\,,
\ee
the charged Higgs masses are
\be
m^2_{H^\pm_i} = -{1\over 2}v^2\left(\Lambda_2  + {\sqrt{3} \over 2}\Lambda_4\right)\quad \mbox{and}
\quad  -{1\over 4}v^2\left(\Lambda_1 + \Lambda_2  + {2 \over \sqrt{3}}\Lambda_4\right)\,.
\ee
The neutral Higgs spectrum contains, as usual, $h$ with mass $m_h^2 = 2 M_0/\sqrt{3}$ and two pairs
of degenerate Higgses with masses
\bea
m^2_{H_i}&=& {v^2 \over 6}\Biggl[\Lambda_3 + {3 \over 2}(\Lambda_1+\Lambda_2) - 4 \tilde \Lambda
\nonumber\\
&& \pm\ \sqrt{\left(\Lambda_3 - {3 \over 2}(\Lambda_1+\Lambda_2) +2 \tilde \Lambda\right)^2 
+ 3\left[(\Lambda_1 - \Lambda_2)^2 + \Lambda_4^2\right]\sin^2(2\alpha-\gamma)}\Biggr]\,.
\eea
In the $S_4$-symmetric case ($\Lambda_4=0$ and $\gamma = 0$), we recover the results of Section~\ref{section-unexpected-S4}.
\end{itemize}
For the first three cases, the Higgs spectra were also explicitly written in \cite{toorop1}.
We have checked that our results fully coincide with theirs.
For the last case, the minimum $(\pm 1, e^{i\pi/3}, e^{-i\pi/3})$ was not explicitly written and analyzed by the authors of \cite{toorop1}. 
Instead they performed a numerical analysis of various minima of the type $(r, e^{i\alpha}, e^{-i\alpha})$,
but their expressions, taken literally, become indeterminate at $r = \pm 1$ and $\alpha = \pi/3$.
It is conceivable that this last point, if treated as a special case, can still be recovered from their starting expressions.

\end{document}